\documentclass[conference,letterpaper]{IEEEtran}

\usepackage{url}
\usepackage{todonotes}
\usepackage{underscore}
\usepackage{cleveref}
\usepackage{tikz}
\usepackage{amsmath}
\usepackage{amssymb}
\usepackage{verbatim}
\usetikzlibrary{positioning}
\usepackage{balance}

\usepackage{tikz}
\usepackage{textcomp}
\usepackage{hyperref}
\usepackage{lipsum}

\newcommand\copyrighttext{%
\footnotesize
This is a preprint of an accepted paper.
  \textcopyright 2017 IEEE. Personal use of this material is permitted.
  Permission from IEEE must be obtained for all other uses, in any current or future
  media, including reprinting/republishing this material for advertising or promotional
  purposes, creating new collective works, for resale or redistribution to servers or
  lists, or reuse of any copyrighted component of this work in other works.
  }
\newcommand\copyrightnotice{%
\begin{tikzpicture}[remember picture,overlay]
\node[anchor=north,yshift=-10pt] at (current page.north) {\fbox{\parbox{\dimexpr\textwidth-\fboxsep-\fboxrule\relax}{\copyrighttext}}};
\end{tikzpicture}%
}

\newtheorem{definition}{\bf Definition}

\title{A Fuzzy Approach to Qualification in Design Exploration for Autonomous Robots and Systems}

\author{\IEEEauthorblockN{Jeremy Morse,\\Dejanira Araiza-Illan,\\and Kerstin Eder}
\IEEEauthorblockA{Dept. of Computer Science\\University of Bristol\\
Bristol, BS8 1UB, UK\\
Email: \{jeremy.morse,dejanira.araizaillan,\\kerstin.eder\}@bristol.ac.uk}
\and
\IEEEauthorblockN{Jonathan Lawry}
\IEEEauthorblockA{Dept. of Engineering Mathematics\\University of Bristol\\
Bristol, BS8 1UB, UK\\
Email: j.lawry@bristol.ac.uk}
\and
\IEEEauthorblockN{Arthur Richards}
\IEEEauthorblockA{Dept. of Aerospace Engineering\\University of Bristol\\
Bristol, BS8 1TR, UK\\
Email: arthur.richards@bristol.ac.uk}}

\begin{document}

\maketitle
\copyrightnotice

\begin{abstract}
Autonomous robots must operate in complex and changing environments subject to requirements on their behaviour.  
Verifying absolute satisfaction (true or false) of these requirements is challenging.  
Instead, we analyse requirements that admit flexible degrees of satisfaction.  
We analyse vague requirements using fuzzy logic, and probabilistic requirements using model checking.  
The resulting analysis method provides a partial ordering of system designs, identifying trade-offs between different requirements in terms of the degrees to which they are satisfied.  
A case study involving a home care robot interacting with a human is used to demonstrate the approach. 
\end{abstract}

\section{Introduction}

If robots are to interact with people routinely, evidence must be provided that they are both safe and function as intended, i.e.\ useful.  
Verification by post-implementation experimenting is expensive, as is correcting any problems found at that stage. 
Hence, verification is preferred at design time, where many aspects of the system, its requirements, and the environment, might not be precisely defined and specified.  
We therefore propose a more flexible form of verification to assess levels of requirement satisfaction at design time, rather than strict pass or failure.  
In particular, we consider two different kinds of requirements: probabilistic requirements (PRs) that formally define requirements which hold with some measurable probability (e.g.\ the robot needs to reach a safe location in 30 seconds with a probability of 0.8); and vague requirements (VRs), i.e.\ requirements that can be specified as a scale between unacceptable and completely acceptable (e.g.\ the velocity of the robot should be preferably low to avoid dangerous collisions).

The contribution of this paper is a single analysis method for requirement satisfaction, combining the use of fuzzy logic for VRs,  and measuring probabilities of satisfaction for PRs.  
These metrics provide a partial ordering of different design candidates that represents a flexible specification.
Every design candidate is evaluated both in terms of fuzzy set membership for every VR and probability for every PR, calculated using probabilistic model checking~\cite{prism}.
Our analysis ensures the levels of VR and PR satisfaction for each candidate are directly comparable.
The combination enables trade-offs between the different kinds of requirements to be investigated during the robot design activity, allowing ranking and discrimination between design candidates.

Consider an autonomous home-care assistant which must periodically attend its user.  
To avoid harmful collisions, it is required that the robot movement be `slow', an example of a VR since the precise threshold of acceptability is to be determined.  
However, it is also required that the robot should reach the user quickly when called, considered as a PR due to the uncertain location of the user.  
Since the latter requires the robot to move quickly, it competes with the `slow' VR.

The presence of multiple requirements has been identified as a serious research challenge for autonomous systems design, e.g.\ when dealing with robots and environments that adapt, due to associated uncertainty and vagueness~\cite{saslandscape}.
The self-adaptive software research roadmap in~\cite{sasroadmap} proposes that requirements be given in \textit{flexible} terms, allowing a broad interpretation of when a system satisfies a requirement and when it does not. 
Thus, formalization and analysis methods for both PRs and VRs combined is timely and necessary to aid designers towards safe and functional robots.

An outline of the remainder of this paper is as follows: In Section~\ref{relatedwork} we give a brief overview of related work.
Section~\ref{prelims} introduces concepts underpinning our work.
Section~\ref{specifications} formalizes a system specification.
Section~\ref{fuzzymodel} proposes how fuzzy logic can be used to represent VRs.
In Section~\ref{casestudy} we describe the care assistant case study and use it to illustrate and evaluate our approach.
Finally, in Section~\ref{conclusion} we conclude and give an outlook on future research.

\section{Related Work}\label{relatedwork}

There is considerable literature on the verification of requirements at design time, i.e.\ before system deployment in the real world. 
Available approaches for a well defined set of requirements include formal methods (e.g.\ model checking in~\cite{webster15toward,Gario2016}), and simulation-based testing (e.g.\ as in~\cite{araizaillan2016}).
The complexity of verifying autonomous systems at design time can be addressed by runtime verification, especially for systems with adaptation capabilities (e.g. \cite{runtimevandv,needsqvruntime}).
This still leaves us with choices to consider at design time, such as the adaptation methodologies, or physical constraints due the choice of sensors and actuators.

Formal methods can be used to verify that PRs are satisfied at design time, e.g.\ probabilistic model checking~\cite{prism}.
Trade-offs in the simultaneous satisfaction of multiple PRs from a multi-objective perspective can be computed (i.e.\ a Pareto front) from solving games with rewards in probabilistic model checking~\cite{Forejt2012}. 
From another perspective, many formal approaches have been proposed for correct-by-construction controller synthesis in robotics and autonomous systems (e.g.\ \cite{Kressgazit2009}). 
Strategies and controllers that partially satisfy a temporal logic property~\cite{Tumova2013}, or that violate a number of assumptions about a problem the least~\cite{Ehlers2014}, can be synthesized. 
Overall, formal approaches at design time are effective for analysing PRs, but do not address VRs and, to date, have not been combined with VR techniques.

For VRs, the literature focuses on quantifying them through metrics (e.g.\ in multi-objective optimization used for design exploration~\cite{Zuluaga2012}).
In the field of requirements engineering, techniques have been proposed for reasoning about trade-offs between requirements that can only be partially satisfied at design time~\cite{partialgoal}, and also techniques to model qualitative VRs through fuzzy logic~\cite{Serrano2011}, albeit at runtime.
Requirement languages such as RELAX~\cite{relax,relaxwithfltl} provide a means for describing and quantifying partial satisfaction of a VR. 
The problem of finding a system configuration that satisfies functional and desirability criteria (VRs) has been formalized by~\cite{requirementsprob}. 
Existing techniques either limit the VRs that can be expressed (such as requiring independence~\cite{partialgoal}) or are not suitable for design time analysis. 
In addition, while most of the approaches offer formal descriptions of trade-offs with VRs, practical techniques to establish if a system satisfies all types of requirements (PRs and VRs) remains a challenge.

\section{Preliminaries}
\label{prelims}

For a completely deterministic system with states $x \in X$ and a transition relation $R \subseteq X \times X$, model checking tools can be used to determine if a requirement, formalized as a temporal logic formula $\phi$, holds from a given start state $x_0 \in X$.
More generally, we are often faced with systems with components that can behave randomly, e.g.\ human behaviour, and which are also non-deterministic or imprecisely defined, e.g.\ where there are only limited design-time assumptions about the robot controller.
Such systems can be modeled as a Markov Decision Process (MDP).
\begin{definition}
A MDP (with labelling function) is a tuple M~=~($X$, $x_0$, $U$, $P$, $\Pi$, $L$) where:
\begin{itemize}
	\item $X$ is a finite set of states;
	\item $x_0 \in X$ is the initial state;
	\item $U$ is a finite set of actions;
	\item $P : X \times U \rightarrow Dist(X)$ is a transition probability function, which maps state-action pairs to a probability distribution over X $Dist(X)$;
	\item $\Pi$ is a finite set of atomic propositions;
	\item $L : X \rightarrow 2^\Pi$ is a labeling function that assigns each state a set of atomic propositions in $\Pi$.
\end{itemize}
\end{definition}

Temporal logic formulae over MDPs can be specified in PCTL \cite{pctl} expressed over the labels of the MDP, which provides operators over probabilities to quantify the probability of a particular formula holding.
The latter enables \textit{quantitative verification}, where one can specify that a system will fulfil a PR with a certain probability.
A formula can be evaluated over the system model to determine the precise probability with which it will hold.
Such assurances, and thus quantitative verification and PRs, are valuable where requirements still ``must'' hold, but where external disturbances may cause them to be violated, such as uncontrolled environments.
An autonomous vehicle, for example, may be capable of fulfilling its requirements, but cannot avoid a violation if another vehicle makes a serious error.

\section{Specifications}\label{specifications}
We now introduce our idea of a specification, to describe variations of system designs, and how existing techniques can be used to determine the level of PR satisfaction of a specification. 

A system is defined by its {\em state} $x \in X$, changed by the enactment of an {\em action} $u\in U$, at each state. 
A \emph{specification} is a two-valued membership function defined over states and actions. It delimits the actions that are allowed in each one of the states, always allowing at least one action per state, to avoid deadlock.
Here we are referring to design constraints such as maximum robot speed, minimum battery charge, time to service the human etc.

\begin{definition}
A specification is a function of the state variables and the actions, $f: X \times U \rightarrow \{0,1\}$, where $f(x,u)=1$ defines an action $u$ that is allowed in the state $x$. 
\end{definition}

Different system design constraints lead to a set of specifications, $F$, to analyse at design time.
There is a natural sense of desirability in which specifications can be weakened or strengthened by, for instance, weakening or strengthening individual design constraints. This can be formalized to define a partial ordering on specifications as follows.

\begin{definition}\label{dfn:weakening}
Given two specifications $f$ and $f^\prime$, we say that $f$ is a \emph{weakening} of $f^\prime$, denoted $f \preceq f^\prime$, if and only if $\forall x,u \in X \times U$, $f(x,u) \geq f^\prime(x,u)$.
\end{definition}

The relation $\geq$ is defined over the two-valued membership set $\{0,1\}$. 
Notice that for a specification $f$, $f^{-1}(1)$ corresponds to the permissible set of pairs of states and actions. Clearly, $f \preceq f^\prime$ if and only if $f^{-1}(1) \supseteq (f^{\prime})^{-1}(1)$.

\subsection{Model Checking a Specification}
\label{mcaspec}

We produce a system representing that specification.
Given an existing system described as an MDP ($X$, $x_0$, $U$, $P$, $\Pi$, $L$), we create a restricted system ($X^\prime$, $x_0$, $U$, $P^\prime$, $\Pi$, $L$), that conforms to a specification $f$. Here,

\begin{equation}
\begin{array}{lcll}
	P'(x, u) & = & P(x, u) & if \  f(x, u) = 1 \\
	P'(x, u) & = & \emptyset & if \  f(x, u) = 0 \\
\end{array}
\end{equation}

\noindent where $\emptyset$ is the empty probability distribution, and $X^\prime$ is the subset of $X$ that is reachable under the transition function $P^\prime$.

Probabilistic model checking tools such as PRISM ~\cite{prism} can be used to determine the probability of a PR $\phi$ holding over an MDP as the system evolves, given the states and actions.
In the presence of non-determinism (i.e.\ multiple actions per state) PRISM can evaluate any Markovian \textit{policy} that resolves multiple actions to one per state.
The evaluation of all policies forms a credal set ${\cal P}_f$, for a given specification $f$.
From this we can determine lower and upper probabilities for $\phi$, from:
\begin{equation}
\begin{array}{l}
\underline{P}_f(\phi)=\max\{P(\phi):P\in {\cal P}_f\}\\
\overline{P}_f(\phi)=\min\{P(\phi):P\in {\cal P}_f\},
\end{array}  
\end{equation}
where $P(\phi)$ is a single probability value for PR $\phi$.
Furthermore, for reasons that are beyond the scope of this paper, it also holds that for probabilistic model checkers if $f \preceq f^\prime$ then ${\cal P}_f \supseteq {\cal P}_{f^\prime}$.
Consequently, the upper probability of the PR $\phi$ increases as the specification is weakened; i.e.\ if $f\preceq f^\prime$ then $\overline{P}_f(\phi) \geq \overline{P}_{f^\prime}(\phi)$.
In the context of design-time verification, upper probabilities are also interesting since they are associated with the use of an optimal controller.
In contrast, lower probabilities are associated with rather unrealistic optimal feedback controllers, such a robot that constantly moves away from its target.
For the rest of this paper, we focus only on the maximum achievable success probability, corresponding to the best possible controller policies.

\section{Fuzzy Modelling of Vaguely Defined Requirements}\label{fuzzymodel}

We propose a formalization of the level of satisfaction of VRs, i.e.\ informal requirements that range from unacceptable to fully acceptable, based on fuzzy logic.
We apply a similar modeling approach to utility theory for multi-agent systems, where utilities are modeled in terms of fuzzy sets and membership functions (e.g.\ in~\cite{Seo}). 
Fuzzy logic allows flexible reasoning with imprecision and uncertainty, also helping to model vagueness of linguistic or intuitive information. We propose that VRs be modeled by normalised fuzzy sets on system actions as follows:

\begin{definition} 
A vague requirement $\chi$ is defined as a fuzzy set with membership function $\mu_{\chi}: f^{-1}(1) \rightarrow [0,1]$ such that $\sup\{\mu_\chi(x,u): (x,u) \in f^{-1}(1) \}=1$. 
\end{definition}

This can then be extended to specifications by quantifying the minimum level to when those state action pairs which satisfy a specification also specify $\chi$.

\begin{definition} \label{def:fuzzycompliance}
Let $f$ be a specification, $\mu_{\chi}(f)$ quantifies the level to which the specification $f$ satisfies the VR $\chi$, given by $\mu_{\chi}(f)=\inf \{\mu_{\chi}(x,u): (x,u) \in f^{-1}(1) \}$.
\end{definition}

Notice that, given this definition and Definition \ref{dfn:weakening}, it follows that if $f \preceq f^\prime$ then $\mu_{\chi}(f) \leq \mu_{\chi}(f^\prime)$.

For separate VRs relating to different aspects of the system, a conjunction of these can be defined as $\chi=\chi_1 \wedge \ldots \wedge \chi_m$, where each VR $\chi_i$ is modeled as a fuzzy set. 
In this case the membership function $\mu_\chi$ is determined using a t-norm $T$ in the standard way so that $\forall (x,u) \in f^{-1}(1), \mu_{\chi}(x,u)=T(\mu_{\chi_1}(x,u), \ldots, \mu_{\chi_m}(x,u))$.

\subsection{A Partially Ordered Set of Specifications}

It is proposed that a finite set of specifications $F$ are considered so as to provide reasonable coverage of the design space. 
Taken together with the weakening ordering in Definition \ref{dfn:weakening} these generate a partially ordered set $(F,\preceq)$ which can be explored on the basis of levels of VR satisfaction.
Moreover, the poset can also be explored on the basis of quantified satisfiability of PRs. 

\subsection{Complementary Use of PR and VR Analyses}

We now bring together the PR evaluation from Section~\ref{mcaspec} and our proposed formalization for VR partial satisfaction.
In order to determine the specification with the highest level of VR satisfaction, amongst those which satisfy the PR $\phi$ to an appropriate probability threshold of $\rho$ (according to quantitative model checking results), we need only consider the least weak of those specifications which satisfy $\phi$.
Given PR $\phi$ and VRs $\chi$, we aim to identify 
\begin{gather*}
\arg \max \{ \mu_{\chi}(f): \overline{P}_f(\phi)\geq \rho,  \ f \in F \}
\end{gather*}
for a specified satisfiability parameter $\rho \in [0,1]$. Since $\overline{P}_f(\phi)$ increases and $ \mu_{\chi}(f)$ decreases as $f$ is weakened, it follows that this is equivalent to
\begin{gather*}
\arg \max \{ \mu_{\chi}(f): f \in W \}
\end{gather*}
where $W=\{f \in F: \overline{P}_f(\phi)\geq \rho\}$. 

The set $W$ can be automatically identified by using search methods over a fully partial ordered set $(F, \preceq)$, although this might be computationally expensive. 
Alternatively, we could sample over $F$ and use the acquired information to find these requirement ``optimal'' specifications at a more efficient computational cost (as is done in Gaussian process regression~\cite{Zuluaga2012}). Overall, the complementary use of fuzzy logic and probabilistic model checking can allow the designer to observe the trade-off between satisfying the different VRs to a high level and satisfying the PRs (e.g.\ to a high probability), for system design exploration.

\section{Case study}\label{casestudy}

A robotic home healthcare assistant needs to assist a human at certain times. 
The living space is laid out as a two-dimensional grid featuring a recharging point, visiting which results in the battery being instantly recharged.
The human's behaviour is assumed to be random, as they occasionally run away from the robot. 

\subsection{Modeling of vague requirements}

There are three safety design constraints; these are the maximum velocity limit $v$, the minimum energy margin before recharge $e$, and the maximum servicing time limit $t$.
Thus, an action $u \in U$ is delimited by a velocity constraint, a battery charge constraint, and a servicing time limit. 
In our model, the design constraints to explore are defined as the sets of ordered constraints $t_{\gamma}= t \leq \gamma$ with $\gamma=1,\ldots,10$ units; $v_{\delta} = v \leq \delta$ with $\delta=1,\ldots,6$ units; and $e_{\zeta}=e \leq \zeta$ with $\zeta=1,\ldots,5,10,15,20,25$ units, respectively for the time, velocity and energy limits. 
By taking conjunctions of these constraints we obtain a set of specifications $F=\{f_{\gamma,\delta,\zeta}:\gamma, \delta, \zeta \}$ where
\begin{eqnarray}\label{myspec}
f_{\gamma,\delta,\zeta} &=& t_{\gamma} \wedge v_{\delta} \wedge e_{\zeta} \ \mathit{where}\ \gamma \in \{1, \ldots, 10\},\\&& \delta \in \{1, \ldots, 6\} \nonumber \ \mathit{and}\ \zeta \in \{1, \ldots, 5,10,15,20,25\}. 
\end{eqnarray}

Hence, the number of varied design constraints is $10+6+9=25$, and $|F|=10\times6\times9=540$. 

\begin{figure*}
\centering
\hspace*{-10mm}
\includegraphics[width=\textwidth]{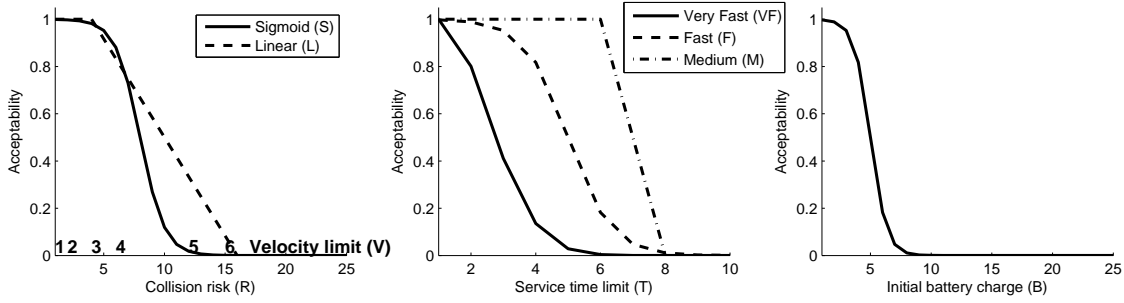}
\caption{Proposed membership functions to quantify partial VR satisfaction of system specifications, with respect to the VRs $\chi_1$, $\chi_2$ and $\chi_3$.\label{fig:differenttimes}}
\vspace{-5mm}
\end{figure*}

Our VRs are as follows, we wish the robot to move at a speed that minimizes the risk of a harmful collision, but to be able to reach the human quickly, while maintaining as low an energy threshold as possible before returning to the charging station.
These are modeled by three fuzzy sets $\chi_1$, $\chi_2$ and $\chi_3$.
The membership functions of these fuzzy sets are shown in Fig.~\ref{fig:differenttimes}.
The overall VR for the system is the conjunction $\chi=\chi_1 \wedge \chi_2 \wedge \chi_3$, where the membership for $\chi$ is determined by applying the $\min$ t-norm:
\begin{gather*}
\mu_\chi(f)=\min( \mu_{\chi_1}(f), \mu_{\chi_2}(f), \mu_{\chi_3}(f) )
\end{gather*}
The choice of membership functions capture the designer's subjective judgement concerning partial requirement satisfaction.
For $\chi_1$, we employ a point system for risk analysis with values from $1$ to $25$, where a low velocity has a lower collision risk (thus a smaller risk value) whereas a high velocity has a higher collision risk (thus a high risk value).
Two possible membership functions are suggested for $\chi_1$.
According to the `Sigmoid' function, a maximum velocity limit below 3 units, with an associated collision risk of below 5 units, is highly desirable; and a maximum velocity limit of 5 units is very undesirable, with a risk of above 10.
In contrast, according to the `Linear' membership function there is more tolerance of collision risk values between 5 and 15, with a steady decay in desirability.
For $\chi_2$, we proposed three different functions that represent a range of service time tolerances, from a robot that has to reach the human as quickly as possible at all times (`Very Fast'), to a robot that can choose to act as fast as possible or slightly slower (`Medium') according to the occasion.
For $\chi_3$, we only proposed one membership function, where an energy threshold of 3 or less is highly desirable, and above 8 is highly undesirable. 

\subsection{Model Checking of a probabilistic requirement}\label{formalanalysis}

To evaluate our PR, we build an MDP model and evaluate it with the probabilistic model checker PRISM~\cite{prism}.
The PR $\phi$ is expressed as a PCTL \cite{pctl} formula:
\begin{equation}
	\phi = \  \neg service \  {\sf U} \  service
\end{equation}
where $service$ is a label that is true for a state where the user is being attended, and false otherwise.
Thus, our PR is that from any start state, the system is able to attend to the user with as high a probability as possible.

Our model is written in the PRISM modeling language, consisting of 5 modules, corresponding to the human and robot motion, the timing, the energy in the battery, and the servicing task. 
The movement of the human is a probabilistic choice of moving (north, south, east, west), or to stand still, each with a probability of 0.2.
The robot and its control system are modeled as a set of non-deterministic choices.
The robot may chose to move at a range of speeds in any of the four directions, or stay still.
We leave the model checker to derive the best navigation policy.
The robot has a battery with a finite amount of energy that, when fully discharged, causes the robot to stop.
We filter out unrealistic robot behaviours, such as moving outside the grid environment, and motions that cause immediate collision with the human.

For each specification $f \in F$ we restrict the model as in Section~\ref{mcaspec} by conjoining propositions with $\phi$, for example $velocity \le 5$, forcing the model checker to not explore any transitions where $velocity$ exceeds 5.
Different time bounds in the specification are modeled by using the ${\sf U} \leq \alpha$ operator instead of ${\sf U}$, specifying that $\phi$ must hold within $\alpha$ steps.
The model checker is used to compute the maximum probability the design can achieve when operating under the specification, assuming an optimal controller, for every start state of the model (i.e.\ combinations of robot, human, battery life etc.).

\subsection{Experimental Results}\label{results}

Experiments were run on a PC with Intel i5-3230M 2.60 GHz CPU, 8 GB of RAM, Ubuntu 14.04. 
We ran the model checking analysis phase on PRISM 4.2.beta1, for each of the 540 specifications. 
Model checking took less than 1 minute for each run, with a minimal computation time of 10 seconds.
All underlying data and models are openly available online\footnote{Available from \url{https://github.com/riveras/homecare}}.

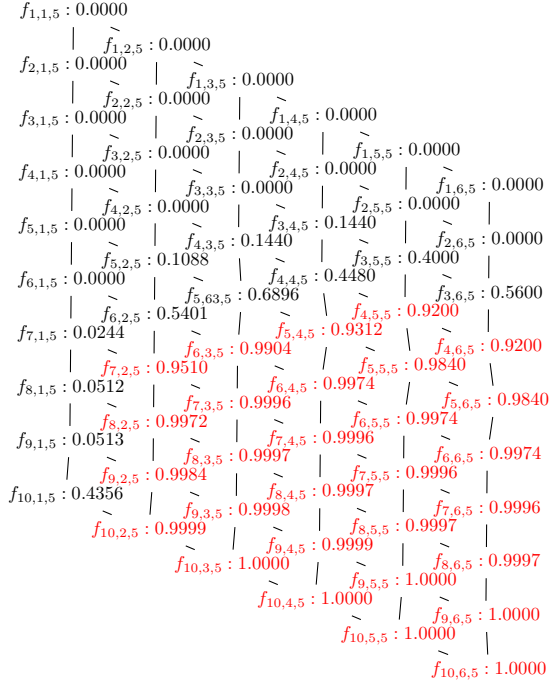
\begin{figure}
\centering
\begin{tikzpicture}[scale=0.7, every node/.style={scale=0.7}]
\node(n1)                          {$f_{1,1,5}: 0.0000$};
\node(n2a)       [below left=0.3cm and -1.6cm of n1] {$f_{2,1,5}:0.0000$};
\node(n2b)      [below right=0.05cm and -0.5cm of n1]  {$f_{1,2,5}: 0.0000$};

\node(n3a)       [below left=0.3cm and -1.6cm of n2a] {$f_{3,1,5}:0.0000$};
\node(n3b)      [below right=0.05cm and -0.5cm of n2a] {$f_{2,2,5}:0.0000$};
\node(n3c)      [below right=0.05cm and -0.5cm of n2b] {$f_{1,3,5}:0.0000$};

\node(n4a)      [below left=0.3cm and -1.6cm of n3a] {$f_{4,1,5}:0.0000$};
\node(n4b)      [below right=0.05cm and -0.5cm of n3a] {$f_{3,2,5}:0.0000$};
\node(n4c)      [below right=0.05cm and -0.5cm of n3b] {$f_{2,3,5}:0.0000$};
\node(n4d)		[below right=0.05cm and -0.5cm of n3c] {$f_{1,4,5}:0.0000$};

\node(n5a)      [below left=0.3cm and -1.6cm of n4a] {$f_{5,1,5}:0.0000$};
\node(n5b)      [below right=0.05cm and -0.5cm of n4a] {$f_{4,2,5}:0.0000$};
\node(n5c)      [below right=0.05cm and -0.5cm of n4b] {$f_{3,3,5}:0.0000$};
\node(n5d)		[below right=0.05cm and -0.5cm of n4c] {$f_{2,4,5}:0.0000$};
\node(n5e)		[below right=0.05cm and -0.5cm of n4d] {$f_{1,5,5}:0.0000$};

\node(n6a)      [below left=0.3cm and -1.6cm of n5a] {$f_{6,1,5}:0.0000$};
\node(n6b)      [below right=0.05cm and -0.5cm of n5a] {$f_{5,2,5}:0.1088$};
\node(n6c)      [below right=0.05cm and -0.5cm of n5b] {$f_{4,3,5}:0.1440$};
\node(n6d)      [below right=0.05cm and -0.5cm of n5c] {$f_{3,4,5}:0.1440$};
\node(n6e)      [below right=0.05cm and -0.5cm of n5d] {$f_{2,5,5}:0.0000$};
\node(n6f)      [below right=0.05cm and -0.5cm of n5e] {$f_{1,6,5}:0.0000$};

\node(n7a)      [below left=0.3cm and -1.6cm of n6a] {$f_{7,1,5}:0.0244$};
\node(n7b)      [below right=0.05cm and -0.5cm of n6a] {$f_{6,2,5}:0.5401$};
\node(n7c)      [below right=0.05cm and -0.5cm of n6b] {$f_{5,63,5}:0.6896$};
\node(n7d)      [below right=0.05cm and -0.5cm of n6c] {$f_{4,4,5}:0.4480$};
\node(n7e)      [below right=0.05cm and -0.5cm of n6d] {$f_{3,5,5}:0.4000$};
\node(n7f)      [below right=0.05cm and -0.5cm of n6e] {$f_{2,6,5}:0.0000$};

\node(n8a)      [below left=0.3cm and -1.6cm of n7a] {$f_{8,1,5}:0.0512$};
\node(n8b)      [below right=0.05cm and -0.5cm of n7a] {\color{red}$f_{7,2,5}:0.9510$};
\node(n8c)      [below right=0.05cm and -0.5cm of n7b] {\color{red}$f_{6,3,5}:0.9904$};
\node(n8d)      [below right=0.05cm and -0.5cm of n7c] {\color{red}$f_{5,4,5}:0.9312$};
\node(n8e)      [below right=0.05cm and -0.5cm of n7d] {\color{red}$f_{4,5,5}:0.9200$};
\node(n8f)      [below right=0.05cm and -0.5cm of n7e] {$f_{3,6,5}:0.5600$};

\node(n9a)      [below left=0.3cm and -1.6cm of n8a] {$f_{9,1,5}:0.0513$};
\node(n9b)      [below right=0.05cm and -0.5cm of n8a] {\color{red}$f_{8,2,5}:0.9972$};
\node(n9c)      [below right=0.05cm and -0.5cm of n8b] {\color{red}$f_{7,3,5}:0.9996$};
\node(n9d)      [below right=0.05cm and -0.5cm of n8c] {\color{red}$f_{6,4,5}:0.9974$};
\node(n9e)      [below right=0.05cm and -0.5cm of n8d] {\color{red}$f_{5,5,5}:0.9840$};
\node(n9f)      [below right=0.05cm and -0.5cm of n8e] {\color{red}$f_{4,6,5}:0.9200$};

\node(n10a)      [below left=0.3cm and -1.6cm of n9a] {$f_{10,1,5}:0.4356$};
\node(n10b)      [below right=0.05cm and -0.5cm of n9a] {\color{red}$f_{9,2,5}:0.9984$};
\node(n10c)      [below right=0.05cm and -0.5cm of n9b] {\color{red}$f_{8,3,5}:0.9997$};
\node(n10d)      [below right=0.05cm and -0.5cm of n9c] {\color{red}$f_{7,4,5}:0.9996$};
\node(n10e)      [below right=0.05cm and -0.5cm of n9d] {\color{red}$f_{6,5,5}:0.9974$};
\node(n10f)      [below right=0.05cm and -0.5cm of n9e] {\color{red}$f_{5,6,5}:0.9840$};

\node(n11b)      [below right=0.05cm and -0.6cm of n10a] {\color{red}$f_{10,2,5}:0.9999$};
\node(n11c)      [below right=0.05cm and -0.5cm of n10b] {\color{red}$f_{9,3,5}:0.9998$};
\node(n11d)      [below right=0.05cm and -0.5cm of n10c] {\color{red}$f_{8,4,5}:0.9997$};
\node(n11e)      [below right=0.05cm and -0.5cm of n10d] {\color{red}$f_{7,5,5}:0.9996$};
\node(n11f)      [below right=0.05cm and -0.5cm of n10e] {\color{red}$f_{6,6,5}:0.9974$};

\node(n12c)      [below right=0.05cm and -0.6cm of n11b] {\color{red}$f_{10,3,5}:1.0000$};
\node(n12d)      [below right=0.05cm and -0.5cm of n11c] {\color{red}$f_{9,4,5}:0.9999$};
\node(n12e)      [below right=0.05cm and -0.5cm of n11d] {\color{red}$f_{8,5,5}:0.9997$};
\node(n12f)      [below right=0.05cm and -0.5cm of n11e] {\color{red}$f_{7,6,5}:0.9996$};

\node(n13d)      [below right=0.05cm and -0.6cm of n12c] {\color{red}$f_{10,4,5}:1.0000$};
\node(n13e)      [below right=0.05cm and -0.5cm of n12d] {\color{red}$f_{9,5,5}:1.0000$};
\node(n13f)      [below right=0.05cm and -0.5cm of n12e] {\color{red}$f_{8,6,5}:0.9997$};

\node(n14e)      [below right=0.05cm and -0.6cm of n13d] {\color{red}$f_{10,5,5}:1.0000$};
\node(n14f)      [below right=0.05cm and -0.5cm of n13e] {\color{red}$f_{9,6,5}:1.0000$};

\node(n15f)		 [below right=0.05cm and -0.5cm of n14e] {\color{red}$f_{10,6,5}:1.0000$};

\draw(n1) -- (n2a);
\draw(n1) -- (n2b);
\draw(n2a) -- (n3a);
\draw(n2a) -- (n3b);
\draw(n2b) -- (n3b);
\draw(n2b) -- (n3c);
\draw(n3a) -- (n4a);
\draw(n3a) -- (n4b);
\draw(n3b) -- (n4b);
\draw(n3b) -- (n4c);
\draw(n3c) -- (n4c);
\draw(n3c) -- (n4d);
\draw(n4a) -- (n5a);
\draw(n4a) -- (n5b);
\draw(n4b) -- (n5b);
\draw(n4b) -- (n5c);
\draw(n4c) -- (n5c);
\draw(n4c) -- (n5d);
\draw(n4d) -- (n5d);
\draw(n4d) -- (n5e);
\draw(n5a) -- (n6a);
\draw(n5a) -- (n6b);
\draw(n5b) -- (n6b);
\draw(n5b) -- (n6c);
\draw(n5c) -- (n6c);
\draw(n5c) -- (n6d);
\draw(n5d) -- (n6d);
\draw(n5d) -- (n6e);
\draw(n5e) -- (n6e);
\draw(n5e) -- (n6f);
\draw(n6a) -- (n7a);
\draw(n6a) -- (n7b);
\draw(n6b) -- (n7b);
\draw(n6b) -- (n7c);
\draw(n6c) -- (n7c);
\draw(n6c) -- (n7d);
\draw(n6d) -- (n7e);
\draw(n6d) -- (n7d);
\draw(n6e) -- (n7e);
\draw(n6e) -- (n7f);
\draw(n6f) -- (n7f);
\draw(n7a) -- (n8a);
\draw(n7a) -- (n8b);
\draw(n7b) -- (n8b);
\draw(n7b) -- (n8c);
\draw(n7c) -- (n8c);
\draw(n7c) -- (n8d);
\draw(n7d) -- (n8d);
\draw(n7d) -- (n8e);
\draw(n7e) -- (n8e);
\draw(n7e) -- (n8f);
\draw(n7f) -- (n8f);
\draw(n8a) -- (n9a);
\draw(n8a) -- (n9b);
\draw(n8b) -- (n9b);
\draw(n8b) -- (n9c);
\draw(n8c) -- (n9c);
\draw(n8c) -- (n9d);
\draw(n8d) -- (n9d);
\draw(n8d) -- (n9e);
\draw(n8e) -- (n9e);
\draw(n8e) -- (n9f);
\draw(n8f) -- (n9f);
\draw(n9a) -- (n10a);
\draw(n9a) -- (n10b);
\draw(n9b) -- (n10b);
\draw(n9b) -- (n10c);
\draw(n9c) -- (n10c);
\draw(n9c) -- (n10d);
\draw(n9d) -- (n10d);
\draw(n9d) -- (n10e);
\draw(n9e) -- (n10e);
\draw(n9e) -- (n10f);
\draw(n9f) -- (n10f);
\draw(n10a) -- (n11b);
\draw(n10b) -- (n11b);
\draw(n10b) -- (n11c);
\draw(n10c) -- (n11c);
\draw(n10c) -- (n11d);
\draw(n10d) -- (n11d);
\draw(n10d) -- (n11e);
\draw(n10e) -- (n11e);
\draw(n10e) -- (n11f);
\draw(n10f) -- (n11f);
\draw(n11b) -- (n12c);
\draw(n11c) -- (n12c);
\draw(n11c) -- (n12d);
\draw(n11d) -- (n12d);
\draw(n11d) -- (n12e);
\draw(n11e) -- (n12e);
\draw(n11e) -- (n12f);
\draw(n11f) -- (n12f);
\draw(n12c) -- (n13d);
\draw(n12d) -- (n13d);
\draw(n12d) -- (n13e);
\draw(n12e) -- (n13e);
\draw(n12e) -- (n13f);
\draw(n12f) -- (n13f);
\draw(n13d) -- (n14e);
\draw(n13e) -- (n14e);
\draw(n13e) -- (n14f);
\draw(n13f) -- (n14f);
\draw(n14e) -- (n15f);
\draw(n14f) -- (n15f);
\end{tikzpicture}
\vspace{-2mm}
\caption{Upper probabilities $\overline{P}_{f_{\gamma,\delta,\zeta}}(\phi)$ of satisfying the PR $\phi$ for a subset of the 540 specifications in $F$ over the form $f_{\gamma,\delta,5}=t_\gamma \wedge v_\delta \wedge e_5$, where  $\gamma \in \{1, \ldots, 10\}$, $\delta \in \{1, \ldots, 6\}$. Specifications with $\overline{P}(\phi) \geq 0.9$ in red. \label{fig:o2d}}
\vspace{-5mm}
\end{figure}

We computed the upper probabilities of satisfying the PR $\phi$, with the system model constrained by each of the specifications $f \in F$ as explained in Section~\ref{formalanalysis}.
Fig.~\ref{fig:o2d} shows a Hasse diagram of the resulting probabilities for a subset of the 540 specifications of the form $f_{\gamma,\delta,5}=t_\gamma \wedge v_\delta \wedge e_5$, where  $\gamma \in \{1, \ldots, 10\}$, $\delta \in \{1, \ldots, 6\}$. 
Fig.~\ref{fig:o2d} also shows the upper probabilities $\overline{P}_f(\phi)$. As described in Section~\ref{mcaspec} these are monotonically increasing when specifications are weakened according to Definition \ref{dfn:weakening}.

The probability bounds provided by $\overline{P}_f(\phi) \geq \rho$ form surfaces analogous to Pareto frontiers in multi-objective optimization, representing the ``optimal'' specifications in terms of meeting the probability threshold $\rho$ for the property $\phi$. 
For example, the strongest specifications such that $\rho = 0.9$ have been highlighted in Fig.~\ref{fig:o2d}.

\begin{figure}
\centering
\begin{tikzpicture}[scale=0.7, every node/.style={scale=0.7}]
\node(n1)                          {$f_{1,1,5}:0.5000$};
\node(n2a)       [below left=0.3cm and -1.6cm of n1] {$f_{2,1,5}:0.5000$};
\node(n2b)      [below right=0.05cm and -0.5cm of n1]  {$f_{1,2,5}: 0.5000$};

\node(n3a)       [below left=0.3cm and -1.6cm of n2a] {$f_{3,1,5}:0.4111$};
\node(n3b)      [below right=0.05cm and -0.5cm of n2a] {$f_{2,2,5}:0.5000$};
\node(n3c)      [below right=0.05cm and -0.5cm of n2b] {$f_{1,3,5}:0.5000$};

\node(n4a)      [below left=0.3cm and -1.6cm of n3a] {$f_{4,1,5}:0.1353$};
\node(n4b)      [below right=0.05cm and -0.5cm of n3a] {$f_{3,2,5}:0.4111$};
\node(n4c)      [below right=0.05cm and -0.5cm of n3b] {$f_{2,3,5}:0.5000$};
\node(n4d)		[below right=0.05cm and -0.5cm of n3c] {$f_{1,4,5}:0.5000$};

\node(n5a)      [below left=0.3cm and -1.6cm of n4a] {$f_{5,1,5}:0.0286$};
\node(n5b)      [below right=0.05cm and -0.5cm of n4a] {$f_{4,2,5}:0.1353$};
\node(n5c)      [below right=0.05cm and -0.5cm of n4b] {$f_{3,3,5}:0.4111$};
\node(n5d)		[below right=0.05cm and -0.5cm of n4c] {$f_{2,4,5}:0.5000$};
\node(n5e)		[below right=0.05cm and -0.5cm of n4d] {$f_{1,5,5}:0.0180$};

\node(n6a)      [below left=0.3cm and -1.6cm of n5a] {$f_{6,1,5}:0.0039$};
\node(n6b)      [below right=0.05cm and -0.5cm of n5a] {$f_{5,2,5}:0.0286$};
\node(n6c)      [below right=0.05cm and -0.5cm of n5b] {$f_{4,3,5}:0.1353$};
\node(n6d)      [below right=0.05cm and -0.5cm of n5c] {$f_{3,4,5}:0.4111$};
\node(n6e)      [below right=0.05cm and -0.5cm of n5d] {$f_{2,5,5}:0.0180$};
\node(n6f)      [below right=0.05cm and -0.5cm of n5e] {$f_{1,6,5}:0.0009$};

\node(n7a)      [below left=0.3cm and -1.6cm of n6a] {$f_{7,1,5}:0.0003$};
\node(n7b)      [below right=0.05cm and -0.5cm of n6a] {$f_{6,2,5}:0.0039$};
\node(n7c)      [below right=0.05cm and -0.5cm of n6b] {$f_{5,3,5}:0.0286$};
\node(n7d)      [below right=0.05cm and -0.5cm of n6c] {$f_{4,4,5}:0.1353$};
\node(n7e)      [below right=0.05cm and -0.5cm of n6d] {$f_{3,5,5}:0.0180$};
\node(n7f)      [below right=0.05cm and -0.5cm of n6e] {$f_{2,6,5}:0.0009$};

\node(n8a)      [below left=0.3cm and -1.6cm of n7a] {$f_{8,1,5}:0.0000$};
\node(n8b)      [below right=0.05cm and -0.5cm of n7a] {\color{red}$f_{7,2,5}:0.0003$};
\node(n8c)      [below right=0.05cm and -0.5cm of n7b] {\color{red}$f_{6,3,5}:0.0039$};
\node(n8d)      [below right=0.05cm and -0.5cm of n7c] {\color{blue}$f_{5,4,5}:0.0286$};
\node(n8e)      [below right=0.05cm and -0.5cm of n7d] {\color{red}$f_{4,5,5}:0.0180$};
\node(n8f)      [below right=0.05cm and -0.5cm of n7e] {$f_{3,6,5}:0.0009$};

\node(n9a)      [below left=0.3cm and -1.6cm of n8a] {$f_{9,1,5}:0.0000$};
\node(n9b)      [below right=0.05cm and -0.5cm of n8a] {\color{red}$f_{8,2,5}:0.0000$};
\node(n9c)      [below right=0.05cm and -0.5cm of n8b] {\color{red}$f_{7,3,5}:0.0003$};
\node(n9d)      [below right=0.05cm and -0.5cm of n8c] {\color{red}$f_{6,4,5}:0.0039$};
\node(n9e)      [below right=0.05cm and -0.5cm of n8d] {\color{red}$f_{5,5,5}:0.0180$};
\node(n9f)      [below right=0.05cm and -0.5cm of n8e] {\color{red}$f_{4,6,5}:0.0009$};

\node(n10a)      [below left=0.3cm and -1.6cm of n9a] {$f_{10,1,5}:0.0000$};
\node(n10b)      [below right=0.05cm and -0.5cm of n9a] {\color{red}$f_{9,2,5}:0.0000$};
\node(n10c)      [below right=0.05cm and -0.5cm of n9b] {\color{red}$f_{8,3,5}:0.0000$};
\node(n10d)      [below right=0.05cm and -0.5cm of n9c] {\color{red}$f_{7,4,5}:0.0003$};
\node(n10e)      [below right=0.05cm and -0.5cm of n9d] {\color{red}$f_{6,5,5}:0.0039$};
\node(n10f)      [below right=0.05cm and -0.5cm of n9e] {\color{red}$f_{5,6,5}:0.0009$};

\node(n11b)      [below right=0.05cm and -0.5cm of n10a] {\color{red}$f_{10,2,5}:0.0000$};
\node(n11c)      [below right=0.05cm and -0.5cm of n10b] {\color{red}$f_{9,3,5}:0.0000$};
\node(n11d)      [below right=0.05cm and -0.5cm of n10c] {\color{red}$f_{8,4,5}:0.0000$};
\node(n11e)      [below right=0.05cm and -0.5cm of n10d] {\color{red}$f_{7,5,5}:0.0003$};
\node(n11f)      [below right=0.05cm and -0.5cm of n10e] {\color{red}$f_{6,6,5}:0.0009$};

\node(n12c)      [below right=0.05cm and -0.6cm of n11b] {\color{red}$f_{10,3,5}:0.0000$};
\node(n12d)      [below right=0.05cm and -0.5cm of n11c] {\color{red}$f_{9,4,5}:0.0000$};
\node(n12e)      [below right=0.05cm and -0.5cm of n11d] {\color{red}$f_{8,5,5}:0.0000$};
\node(n12f)      [below right=0.05cm and -0.5cm of n11e] {\color{red}$f_{7,6,5}:0.0003$};

\node(n13d)      [below right=0.05cm and -0.6cm of n12c] {\color{red}$f_{10,4,5}:0.0000$};
\node(n13e)      [below right=0.05cm and -0.5cm of n12d] {\color{red}$f_{9,5,5}:0.0000$};
\node(n13f)      [below right=0.05cm and -0.5cm of n12e] {\color{red}$f_{8,6,5}:0.0000$};

\node(n14e)      [below right=0.05cm and -0.6cm of n13d] {\color{red}$f_{10,5,5}:0.0000$};
\node(n14f)      [below right=0.05cm and -0.5cm of n13e] {\color{red}$f_{9,6,5}:0.0000$};

\node(n15f)		 [below right=0.05cm and -0.6cm of n14e] {\color{red} $f_{10,6,5}:0.0000$};

\draw(n1) -- (n2a);
\draw(n1) -- (n2b);
\draw(n2a) -- (n3a);
\draw(n2a) -- (n3b);
\draw(n2b) -- (n3b);
\draw(n2b) -- (n3c);
\draw(n3a) -- (n4a);
\draw(n3a) -- (n4b);
\draw(n3b) -- (n4b);
\draw(n3b) -- (n4c);
\draw(n3c) -- (n4c);
\draw(n3c) -- (n4d);
\draw(n4a) -- (n5a);
\draw(n4a) -- (n5b);
\draw(n4b) -- (n5b);
\draw(n4b) -- (n5c);
\draw(n4c) -- (n5c);
\draw(n4c) -- (n5d);
\draw(n4d) -- (n5d);
\draw(n4d) -- (n5e);
\draw(n5a) -- (n6a);
\draw(n5a) -- (n6b);
\draw(n5b) -- (n6b);
\draw(n5b) -- (n6c);
\draw(n5c) -- (n6c);
\draw(n5c) -- (n6d);
\draw(n5d) -- (n6d);
\draw(n5d) -- (n6e);
\draw(n5e) -- (n6e);
\draw(n5e) -- (n6f);
\draw(n6a) -- (n7a);
\draw(n6a) -- (n7b);
\draw(n6b) -- (n7b);
\draw(n6b) -- (n7c);
\draw(n6c) -- (n7c);
\draw(n6c) -- (n7d);
\draw(n6d) -- (n7e);
\draw(n6d) -- (n7d);
\draw(n6e) -- (n7e);
\draw(n6e) -- (n7f);
\draw(n6f) -- (n7f);
\draw(n7a) -- (n8a);
\draw(n7a) -- (n8b);
\draw(n7b) -- (n8b);
\draw(n7b) -- (n8c);
\draw(n7c) -- (n8c);
\draw(n7c) -- (n8d);
\draw(n7d) -- (n8d);
\draw(n7d) -- (n8e);
\draw(n7e) -- (n8e);
\draw(n7e) -- (n8f);
\draw(n7f) -- (n8f);
\draw(n8a) -- (n9a);
\draw(n8a) -- (n9b);
\draw(n8b) -- (n9b);
\draw(n8b) -- (n9c);
\draw(n8c) -- (n9c);
\draw(n8c) -- (n9d);
\draw(n8d) -- (n9d);
\draw(n8d) -- (n9e);
\draw(n8e) -- (n9e);
\draw(n8e) -- (n9f);
\draw(n8f) -- (n9f);
\draw(n9a) -- (n10a);
\draw(n9a) -- (n10b);
\draw(n9b) -- (n10b);
\draw(n9b) -- (n10c);
\draw(n9c) -- (n10c);
\draw(n9c) -- (n10d);
\draw(n9d) -- (n10d);
\draw(n9d) -- (n10e);
\draw(n9e) -- (n10e);
\draw(n9e) -- (n10f);
\draw(n9f) -- (n10f);
\draw(n10a) -- (n11b);
\draw(n10b) -- (n11b);
\draw(n10b) -- (n11c);
\draw(n10c) -- (n11c);
\draw(n10c) -- (n11d);
\draw(n10d) -- (n11d);
\draw(n10d) -- (n11e);
\draw(n10e) -- (n11e);
\draw(n10e) -- (n11f);
\draw(n10f) -- (n11f);
\draw(n11b) -- (n12c);
\draw(n11c) -- (n12c);
\draw(n11c) -- (n12d);
\draw(n11d) -- (n12d);
\draw(n11d) -- (n12e);
\draw(n11e) -- (n12e);
\draw(n11e) -- (n12f);
\draw(n11f) -- (n12f);
\draw(n12c) -- (n13d);
\draw(n12d) -- (n13d);
\draw(n12d) -- (n13e);
\draw(n12e) -- (n13e);
\draw(n12e) -- (n13f);
\draw(n12f) -- (n13f);
\draw(n13d) -- (n14e);
\draw(n13e) -- (n14e);
\draw(n13e) -- (n14f);
\draw(n13f) -- (n14f);
\draw(n14e) -- (n15f);
\draw(n14f) -- (n15f);
\end{tikzpicture}
\vspace{-2mm}
\caption{The membership values for $\chi$ for the same subset of specifications as shown Fig.~\ref{fig:o2d}, using the `Sigmoid' function as $\mu_{\chi_1}$, and the `Very Fast' function as $\mu_{\chi_2}$. Specifications with $\overline{P}_f(\phi) \geq 0.9$ are shown in red and blue. The specification satisfying $\arg \max \{ \mu_{\chi}(f): f \in W \}$ within the subset is shown in blue. \label{fig:tnorm2d}}
\vspace{-5mm}
\end{figure}
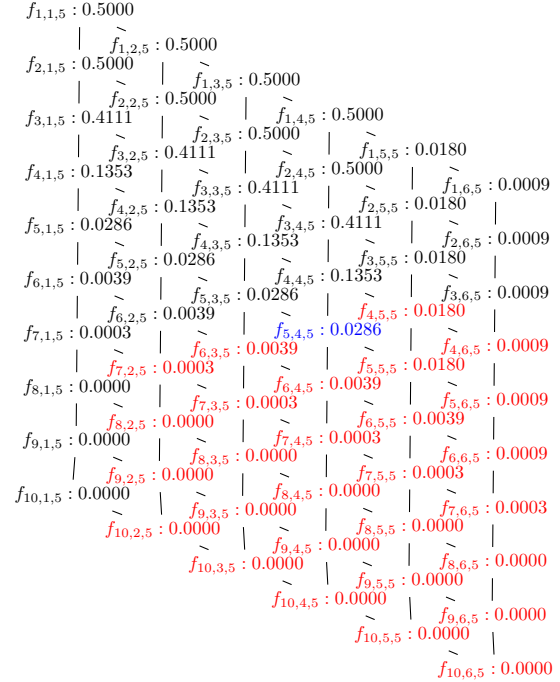

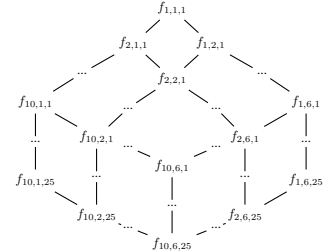
\begin{figure}[t]
\centering
\begin{tikzpicture}[scale=0.6, every node/.style={scale=0.5}]
\node(n1)                           {$f_{1,1,1}$};
\node(n2a)       [below left=0.17cm and 0cm of n1] {$f_{2,1,1}$};
\node(n2b)      [below right=0.17cm and 0cm of n1]  {$f_{1,2,1}$};

\node(n3a)       [below left=0.17cm and 0.3cm of n2a] {...};
\node(n3b)      [below right=0.17cm and 0cm of n2a] {$f_{2,2,1}$};
\node(n3c)      [below right=0.17cm and 0.3cm of n2b] {...};

\node(n4a)      [below left=0.17cm and 0.2cm of n3a] {$f_{10,1,1}$};
\node(n4b)      [below left=0.17cm and 0.2cm of n3b] {...};
\node(n4c)      [below right=0.17cm and 0.2cm of n3b] {...};
\node(n4d)		[below right=0.17cm and 0.2cm of n3c] {$f_{1,6,1}$};
\node(an4a)      [below=0.3cm of n4a] {...};
\node(bn4a)      [below=0.3cm of an4a] {$f_{10,1,25}$};
\node(an4d)      [below=0.3cm of n4d] {...};
\node(bn4d)      [below=0.3cm of an4d] {$f_{1,6,25}$};

\node(n5a)      [below left=0.17cm and 0cm of n4b] {$f_{10,2,1}$};
\node(n5b)      [below=0.34cm of n3b] {};
\node(n5c)      [below right=0.17cm and 0cm of n4c] {$f_{2,6,1}$};
\node(an5a)      [below=0.3cm of n5a] {...};
\node(bn5a)      [below=0.3cm of an5a] {$f_{10,2,25}$};
\node(an5c)      [below=0.3cm of n5c] {...};
\node(bn5c)      [below=0.3cm of an5c] {$f_{2,6,25}$};

\node(n6a)      [below =0.37cm of n4b] {...};
\node(n6b)      [below =0.37cm of n4c] {...};
\node(an6a)      [below=0.3cm of n6a] {};
\node(an6b)      [below=0.3cm of n6b] {};
\node(bn6a)      [below=0.5cm of an6a] {...};
\node(bn6b)      [below=0.5cm of an6b] {...};

\node(n7a)      [below =0.4cm of n5b] {$f_{10,6,1}$};
\node(an7a)      [below=0.3cm of n7a] {...};
\node(bn7a)      [below=0.3cm of an7a] {$f_{10,6,25}$};

\draw(n1) -- (n2a);
\draw(n1) -- (n2b);
\draw(n2a) -- (n3a);
\draw(n2a) -- (n3b);
\draw(n2b) -- (n3b);
\draw(n2b) -- (n3c);
\draw(n3a) -- (n4a);
\draw(n3b) -- (n4b);
\draw(n3b) -- (n4c);
\draw(n3c) -- (n4d);
\draw(n4a) -- (n5a);
\draw(n4a) -- (an4a);
\draw(n4b) -- (n5a);
\draw(n4c) -- (n5c);
\draw(n4d) -- (n5c);
\draw(an4a) -- (bn4a);
\draw(n4d) -- (an4d);
\draw(an4d) -- (bn4d);
\draw(n5a) -- (an5a);
\draw(n5c) -- (n6b);
\draw(an5a) -- (bn5a);
\draw(bn4a) -- (bn5a);
\draw(bn4d) -- (bn5c);
\draw(n5c) -- (an5c);
\draw(an5c) -- (bn5c);
\draw(n5a) -- (n6a);
\draw(n6a) -- (n7a);
\draw(n6b) -- (n7a);
\draw(n7a) -- (an7a);
\draw(an7a) -- (bn7a);
\draw(bn6a) -- (bn7a);
\draw(bn6b) -- (bn7a);
\draw(bn5a) -- (bn6a);
\draw(bn5c) -- (bn6b);
\end{tikzpicture}
\vspace{-3mm}
\caption{Hasse diagram of the specifications in the case study, ordered according to their VR compliance from highest (top) to lowest (bottom), servicing time limits on the right side, and velocity limits on the right side.\label{fig:lattice3d}}
\vspace{-5mm}
\end{figure}

We computed the membership value $\mu_\chi(f)$ for each specification, to quantify the level of VR partial satisfaction.
Fig.~\ref{fig:tnorm2d} shows an example of this computation, for the same specifications as Fig.~\ref{fig:o2d}, using the `Sigmoid' function as $\mu_{\chi_1}$ and the `Very Fast' function for $\mu_{\chi_2}$ from Fig.~\ref{fig:differenttimes}.
The results in Fig.~\ref{fig:tnorm2d} show $\mu_{\chi}(f_{\gamma,\delta,\zeta})$, in the same order as Fig.~\ref{fig:o2d}.
We have highlighted the specifications with a probability $\overline{P}_f(\phi) \geq 0.9$ as in Fig.~\ref{fig:o2d}. 
The specifications that satisfy $\arg \max \{ \mu_{\chi}(f): f \in W \}$ are shown in blue.
Fig.~\ref{fig:lattice3d} shows a compressed version of the $10 \times 6 \times 9$ Hasse diagram with the strongest specification as the top element.

Adopting the `Sigmoid' function for $\mu_{\chi_1}$ and the `Very Fast' function for $\mu_{\chi_2}$ we have that
\begin{gather*}
\arg \max \{ \mu_{\chi}(f): f \in W \} =\{f_{5,4,5}, f_{5,4,4}\},
\end{gather*} 
which conform to the VRs to degree $\mu_{\chi}(f)=0.0286$. 
When using the `Medium' function for $\mu_{\chi_2}$ so that we have a larger maximum servicing time threshold and a lower maximum velocity threshold than above, then 
\begin{gather*}
\arg \max \{ \mu_{\chi}(f): f \in W \} =\{f_{5,4,4}, f_{6,3,4}, f_{6,4,4}\}
\end{gather*}
with $\mu_{\chi}(f)=0.8176$.

These ``optimal'' specifications trade-off an increased risk of collision and a slower service time  so as to meet the PR $\phi$. 
The designer can also see the impact of the energy margins, where thresholds above $5$ have an identical effect on the probabilities.

For the remaining combinations of membership functions, the computed ``optimal'' specifications were: when using the `Sigmoid' function for $\mu_{\chi_1}$ and the `Fast' function for $\mu_{\chi_2}$ and when using the `Linear' function for $\mu_{\chi_1}$ and the `Fast' function for $\mu_{\chi_2}$ we have that, with $\mu_{\chi}(f)=0.5$,
\begin{gather*}
\arg \max \{ \mu_{\chi}(f): f \in W \} =\{f_{5,4,5},f_{5,4,4}\}.
\end{gather*} 

If we use the `Linear' function for $\mu_{\chi_1}$ and the `Very Fast' function for $\mu_{\chi_2}$ then, with $\mu_{\chi}(f)= 0.1353$,
\begin{gather*}
\arg \max \{ \mu_{\chi}(f): f \in W \} =\{f_{4,5,4},f_{4,5,5}\}.
\end{gather*}

Finally, when we use the `Linear' function for $\mu_{\chi_1}$ and the `Medium' function for $\mu_{\chi_2}$ then, with $\mu_{\chi}(f)= 0.8176$,
\begin{gather*}
\arg \max \{ \mu_{\chi}(f): f \in W \} =\{f_{5,4,4}, f_{6,3,4},f_{6,4,4}\}.
\end{gather*}

The differences between choosing a robot that needs to service the human urgently or not, i.e.\ using the `Fast' or `Very Fast', versus the `Medium' function for $\mu_{\chi_2}$, reflect the designer judgements, as the maximum service time threshold is allowed to increase.
The desired urgency also influences the velocity threshold, since increasing the maximum velocity limit decreases the servicing time, at a higher collision risk.

\section{Conclusions and Future Work}\label{conclusion}

In this work, we have considered the challenge of exploring different system designs in the presence of multiple requirements of different forms.
We proposed a formalization of levels of VR satisfaction using fuzzy logic, to characterize and rank different design options which complements traditional PR formal analysis. 
The complementary use of these analyses enhances the designers' understanding of their systems, and of the different design options available to them.

We have used a home care assistant case study to illustrate our approach, exploring variations of design constraints.
We used the probabilistic model checker PRISM to evaluate the upper probability of PRs holding over a formal model of each system design acting in its environment.  
We analysed the VRs through a fuzzy logic formalization. 
A partially ordered set of specifications was generated, and designs that satisfy PRs and VRs to the highest levels possible were identified.

Our technique does not prescribe an algorithm for interpreting or searching
the designs, leaving that to the designer.
However the number of design candidates can increase exponentially with model
detail, similarly to the state explosion problem of model checking.
The level of detail required is again up to the designer, and in many
circumstances it may be that the expense of analysing the large number of
designs is outweighed by the potentially huge cost of adopting an erroneous
design, e.g. harm or financial loss.

Future work will apply our approach to more complex designs for which we will need to investigate more computationally efficient methods of generating and exploring the ordered set of specifications.
For example, we could analyse only some specifications with respect to PR satisfaction (with model checking), to pre-select a set of specifications for VR analysis, reducing the partially ordered set size.

\subsubsection*{Acknowledgement.}
This work was supported by the EPSRC grants EP/J01205X/1 RIVERAS: Robust Integrated Verification of Autonomous Systems, and EP/K006320/1 and EP/K006223/1: Trustworthy Robotic Assistants. 

\balance

\bibliographystyle{IEEEtran}
\bibliography{fm2016}

\end{document}